\documentclass[a4paper]{interact}

\usepackage{color}
\usepackage{enumerate}
\usepackage{listings}

\usepackage[caption=false]{subfig}

\usepackage[numbers,sort&compress]{natbib}
\bibpunct[, ]{[}{]}{,}{n}{,}{,}
\renewcommand\bibfont{\fontsize{10}{12}\selectfont}
\makeatletter
\def\NAT@def@citea{\def\@citea{\NAT@separator}}
\makeatother

\theoremstyle{plain}

\theoremstyle{definition}

\theoremstyle{remark}

\usepackage{braket}

\lstdefinelanguage{json}{
    basicstyle=\normalfont\ttfamily,
    numberstyle=\scriptsize,
    stepnumber=1,
    numbersep=8pt,
    showstringspaces=false,
    breaklines=true,
    frame=lines,
}

\begin{document}

\articletype{ARTICLE TEMPLATE}

\title{Rapid and Robust construction of an ML-ready peak feature table from X-ray diffraction data using Bayesian peak-top fitting}

\author{
    \name{
    Ryo Murakami\textsuperscript{a}, Taisuke T. Sasaki\textsuperscript{b}, Hideki Yoshikawa\textsuperscript{a, c}, Yoshitaka Matsushita\textsuperscript{a}, Keitaro Sodeyama\textsuperscript{c}, Tadakatsu Ohkubo\textsuperscript{b}, Hiroshi Shinotsuka\textsuperscript{a}, Kenji Nagata\textsuperscript{c}
        \thanks{
            CONTACT Kenji Nagata. Email: NAGATA.Kenji@nims.go.jp
        }
    }
    \affil{
        \textsuperscript{a}Research Network and Facility Services Division, National Institute for Materials Science, Tsukuba 305-0044, Japan,\\
        \textsuperscript{b}Research Center for Magnetic and Spintronic Materials, National Institute for Materials Science, Tsukuba 305-0047, Japan,\\
        \textsuperscript{c}Center for Basic Research on Materials, National Institute for Materials Science, Tsukuba 305-0044, Japan,
    }
}

\maketitle

\begin{abstract}
To advance the development of materials through data-driven scientific methods, appropriate methods for building machine learning (ML)-ready feature tables from measured and computed data must be established. In materials development, X-ray diffraction (XRD) is an effective technique for analysing crystal structures and other microstructural features that have information that can explain material properties. Therefore, the fully automated extraction of peak features from XRD data without the bias of an analyst is a significant challenge. This study aimed to establish an efficient and robust approach for constructing peak feature tables that follow ML standards (ML-ready) from XRD data. We challenge peak feature extraction in the situation where only the peak function profile is known a priori, without knowledge of the measurement material or crystal structure factor. We utilized Bayesian estimation to extract peak features from XRD data and subsequently performed Bayesian regression analysis with feature selection to predict the material property. The proposed method focused only on the tops of peaks within localized regions of interest (ROIs) and extracted peak features quickly and accurately. This process facilitated the rapid extracting of major peak features from the XRD data and the construction of an ML-ready feature table. We then applied Bayesian linear regression to the maximum energy product $(BH)_{max}$, using the extracted peak features as the explanatory variable. The outcomes yielded reasonable and robust regression results. Thus, the findings of this study indicated that \textit{004} peak height and area were important features for predicting $(BH)_{max}$.
\end{abstract}

\begin{keywords}
Spectral Decomposition, Bayesian estimation, feature selection, AI-ready
\end{keywords}

\section{Introduction}
\quad In an effort to accelerate material development via data-driven science, the construction of a machine learning (ML)-ready feature table from measurement or computation data is essential. This necessitates appropriate analysis tools that extract the interpretable low-dimensional features from high-dimensional data such as spectra and images. In materials development, X-ray diffraction (XRD) is an effective technique for analysing crystal structures and other microstructural features that have information that can explain material properties. The crystal structure of a material can be understood by analyzing the diffraction peaks in the XRD data. However, XRD data often has numerous peaks and observation points, making it difficult to fully automatically extract the peak features from XRD data without the bias of an analyst.

The peak search method employing differential operation is often used in XRD analysis \cite{1973peakDetermination, 1988peakDetermination, 1993structureDetermination}. Peak search can find the primary peaks; however, it is unable to find sub-peaks, i.e., the peak of a hem. Consequently, the analyst has to manually assign sub-peaks to ensure that an ML-ready peak feature table is not constructed rather rapidly. The measurement data can be used as a feature table; however, it requires strong constraints, such as the need to used identical measurement equipment and measurement conditions. To share the feature table across a variety of material development processes, we believe that the features of the peak that are robust with respect to the measurement equipment and conditions are required. Thus, there is a need for a fully automated peak separation method in case of XRD data.

In recent years, XRD data analysis methods have been proposed for the automatic extraction of material information \cite{yuta2020,yuta2022reviw,2023ReviewXRD,2023XRDanalysis}. These methods are remarkably effective for XRD analysis of well-known materials and experimental systems. However, they require an organized database or a large dataset. In addition, they strongly reduce the XRD data to the lattice parameters. This reduction is stronger than the peak extraction. As a result, the peak parameters containing rich material information cannot be stored in the material database or utilized appropriately. Studies are actively attempting to extract information from large datasets in XRD data using multivariate analysis based on large-scale or non-parametric models \cite{yuta2022,xie2018,ziletti2018,choudhary2021,kovacs2023}. It is desirable to have a large dataset with the same measurement conditions: slit size, scan step, X-ray resolution, etc. However, such a methodology may not be suitable for efficient sharing of data and features from multiple locations/institutions.

This study aimed to develop a method for automatically extracting high-intensity peak features. We challenge peak feature extraction in the situation where only the peak function profile is known a priori, without knowledge of the measurement material or crystal structure factor. This study demonstrated the effectiveness of the proposed framework in predicting the maximum energy product, $(BH)_{max}$, an important magnetic property in permanent magnets such as neodymium. Our framework first extracted the region of interest (ROI) for predicting the material property using partial least squares (PLS) regression \cite{PLS}. The PLS method can perform multivariate analysis for predicting the objective variables and hence can visualize the important regions of the spectrum. Then the proposed method extracted the peak features in the ROIs. The proposed peak separation method extracted peak features using information from the peak tops only. This facilitated the removal of the bias in peaks with a small base. The framework rapidly provided the peak feature from the XRD data by focusing on the peak top in an ROI. Furthermore, this method was used to calculate the accuracy of peak feature extraction using Bayesian estimation \cite{2012BayesianSpectral,2016UseBayesian, 2017UseBayesian,murakami2023}.

In this study, $(BH)_{max}$ was regressed using the peak features based on the Bayesian linear regression model with the feature selection \cite{2022BMAobinata}. This regression model can estimate the probability of a feature being used in the regression by pseudo-exploring all combinations using the replica exchange Monte Carlo method \cite{remc,2012BayesianSpectral}. Consequently, the proposed framework indicated that the peak area and peak height of the \textit{004} plane peak were important to predict the $(BH)_{max}$. Therefore, our framework can provide highly interpretable analysis results by using peak features. 

\section{Data}\label{sec:data}
\subsection{X-ray diffraction data of Neodymium magnets by hot extrusion}\label{sec:xrd}
\begin{figure}
    \centering
    \includegraphics[width=\linewidth]{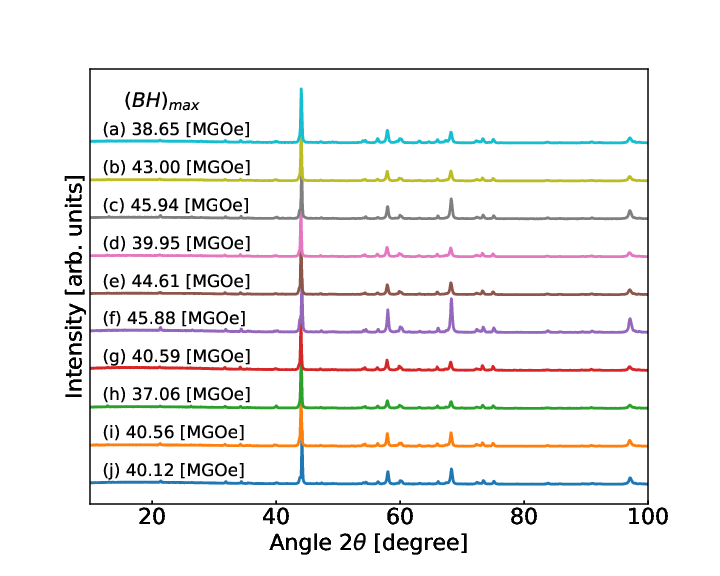}
    \caption{Part of measured X-ray diffraction (XRD) datasets of $\mathrm{ Nd_{2}Fe_{14}B }$ magnets under various hot extrusion conditions. The hot extrusion temperature $T_\mathrm{ext}$ [${}^\circ$C] and load limit $F_\mathrm{ext}$ [kN] for each data are shown as follows: [ (a) $T_\mathrm{ext}=750$, $F_\mathrm{ext}=70$, (b) $T_\mathrm{ext}=750$, $F_\mathrm{ext}=100$, (c) $T_\mathrm{ext}=750$, $F_\mathrm{ext}=60$, (d) $T_\mathrm{ext}=750$, $F_\mathrm{ext}=60$, (e) $T_\mathrm{ext}=750$, $F_\mathrm{ext}=50$, (f) $T_\mathrm{ext}=775$, $F_\mathrm{ext}=50$, (g) $T_\mathrm{ext}=775$, $F_\mathrm{ext}=50$, (h) $T_\mathrm{ext}=750$, $F_\mathrm{ext}=50$, (i) $T_\mathrm{ext}=775$, $F_\mathrm{ext}=50$, (j) $T_\mathrm{ext}=750$, $F_\mathrm{ext}=35$ ]. }
    \label{fig:raw_data}
\end{figure}

\quad We measured the XRD and magnetic properties, such as the maximum energy product $(BH)_{max}$, of Neodymium magnets fabricated by hot extrusion. In the dataset denoted as $\{ (\mathcal{D}_m, t_m)\}_{m=1}^{M}$, $M$ is the number of data samples. In this paper, an XRD datapoint is denoted as $\mathcal{D}=\{(\bm{x}, \bm{y})\}=\{(x_n, y_n)\}_{n=1}^{N}$ where $N$ is the number of data points, and a data point $(x_n, y_n)$ indicates the observation angle $2\theta$ [degree] and intensity [counts], respectively. Figure \ref{fig:raw_data} shows a part of the measured XRD data. The symbol $\bm{t}=(t_1, t_2, \ldots, t_M)^{\top}$ denotes the magnetic properties, for example, the maximum energy product $(BH)_{max}$.

In this measurement data, the number of XRD data was 176, and one XRD dataset had 9000 data points. The range of the observation angle $2\theta$ was 10.0–100.0 [degree], and the observation step was 0.01 [degree]. The intensity unit of XRD datapoints was the counting unit [counts].

\section{Concept}\label{sec:concept}
\quad The concept of our framework  is to quickly formulate of the ML-ready feature table from XRD data. In particular, we focused on the peak features to achieve high interpretability. To realize the rapid construction of feature tables, the proposed method extracted peak features for ROIs with large peak intensities. This method estimated the posterior distribution of the peak features by extracting the peaks through Bayesian estimation using the replica exchange Monte Carlo method. This facilitated discussions on the estimation accuracy of the extracted peak features. Furthermore, the method provided analyst-independent global solutions so that the extracted peak features were less analyst-dependent. In addition, we regressed a material property on the peak features using Bayesian linear regression with feature selection. Thus, the proposed method provides peak extraction and regression analysis based on a full Bayesian framework.

\section{Result and discussion}\label{sec:result}
\begin{figure}
    \centering
    \includegraphics[width=\linewidth]{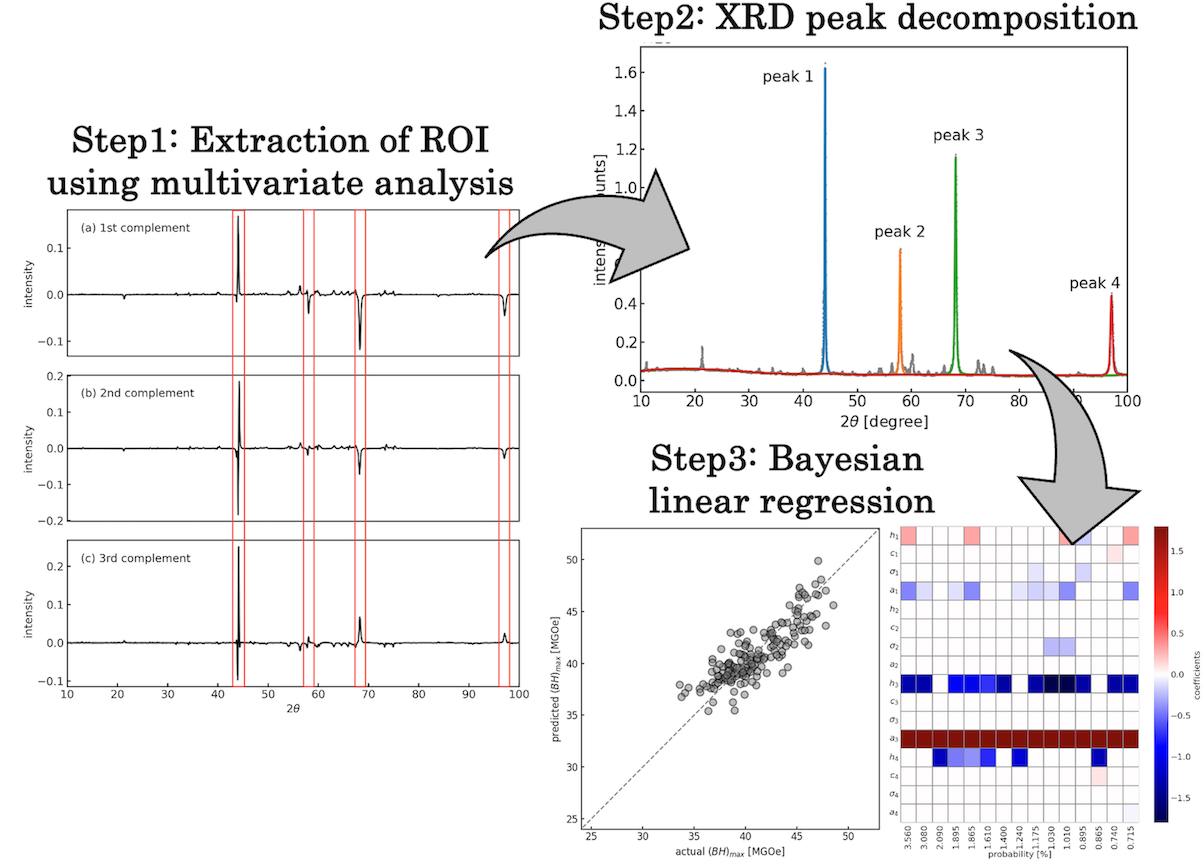}
    \caption{Analysis workflow in our framework}
    \label{fig:result:workflow}
\end{figure}

\quad In this study, we constructed an ML-ready peak feature table from XRD data following three steps. Figure \ref{fig:result:workflow} shows the schematic of the three steps: (Step 1) ROI extraction using PLS regression, (Step 2) XRD peak decomposition using Bayesian estimation, (Step 3) Bayesian linear regression with feature selection. This section discusses the results of each step.

\subsection{Extraction of ROI using the PLS regression}\label{sec:roi}
\quad The proposed framework first extracted the ROI from XRD data using PLS regression. PLS regression reduces the dimensionality of the input data to latent variables based from which the objective variable (material properties) can be predicted. We decided that the ROIs should be those with information that could predict the objective variable.

\begin{figure}
    \centering
    \includegraphics[width=\linewidth]{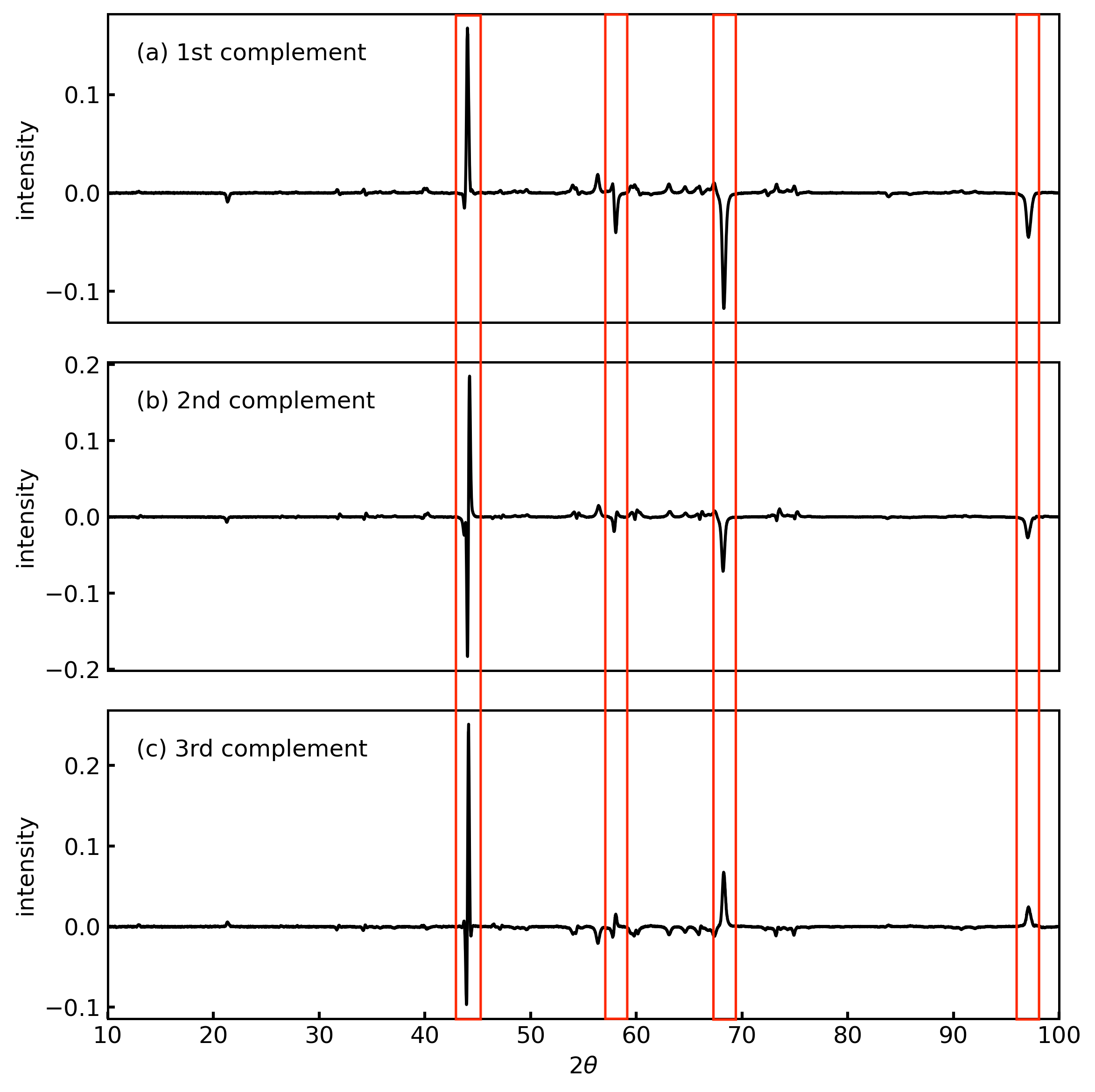}
    \caption{Basis data obtained by PLS regression}
    \label{fig:result:pls-ref-weights}
\end{figure}

Figure \ref{fig:result:pls-ref-weights} shows the basis components obtained by PLS regression. A correlation with the objective variable was observed when setting to three components. We present the results of the PLS regression when the number of components is set to three. Figure \ref{fig:result:pls-ref-weights} (a)--(c) show the 1st, 2nd, and 3rd components, respectively. As shown in Figure \ref{fig:result:pls-ref-weights}, there were approximately four diffraction angle regions (ROI) of high component intensity, as indicated by the red border. This study focused on the four ROIs and performed the XRD peak decomposition. The result of peak decomposition is presented in sub-section \ref{sec:peak}.

\subsection{Peak feature extraction using peak-top fitting}\label{sec:peak}
\quad Extracting peak features is crucial for the understanding of humans. This sub-section presents the results of XRD peak fitting. The peak decomposition model is described  in Appendix \ref{sec:model:peak}. Appendix \ref{sec:model:peak} describes the setting of the prior distributions in the Bayesian estimation of the peak features. In the defined model, we set the intensity spread $\sigma_h$, the position spread $\sigma_c$, width mode $\mu_{\sigma}$, and width spread $\sigma_{\sigma}$ as the hyper-parameters of the prior distribution. We set the hyper-parameters of the prior distribution to $\sigma_h=2.0$, $\sigma_c = 1.0$, $\mu_{\sigma} = 0.1$, $\sigma_{\sigma} = 0.2$. Our method extracted peak features focusing on the four ROIs. We performed peak fitting by assuming the existence of one dominant main peak in the ROI for improved interpretability and high calculation throughput. Thus, the fine structure of the peak hem was not considered in this study.

However, if the minor peaks cannot be assigned, a bias is observed in the estimation of the major peaks. Therefore, in this study, a method was developed to automatically extract the major peak features. In this case, the bias of the minor peaks was removed by using information related only to the peak tops. 

\begin{figure}
    \centering
    \includegraphics[width=\linewidth]{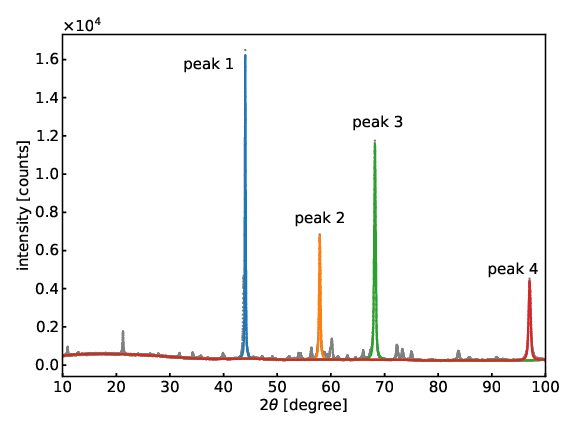}
    \caption{One example of XRD peak fitting focused on peak top in ROIs}
    \label{fig:exsample_fitting}
\end{figure}

Figure \ref{fig:exsample_fitting} presents an example of the peak fitting result of our method. In this study, we performed peak fitting using 30 points around the peak top. In this figure, the gray scatter plot indicates all data points, and the blue, orange, green, and red peaks are the fitted peaks. We indexed the peaks in order from the low-angle side. As shown in Figure \ref{fig:exsample_fitting}, the extracted four peaks well represented the characteristic of high-intensity diffraction peaks. We describe the posterior distribution of this fitting in the Appendix \ref{sec:sup:posterior}. Our method transformed 9000 observation data points into 12 parameters $\Theta=\{h_k, c_k, \sigma_k | k=1,\ldots,\mathrm{K}=4\}$ by peak fitting. We extracted the peak features exhibiting a Bayesian posterior distribution from 176 XRD datapoints in approximately 180 min, despite the sequential analysis of XRD data. Therefore, our method facilitated obtaining peak features with confidence intervals in approximately 1 min.

\begin{figure}
    \centering
    \includegraphics[width=\linewidth]{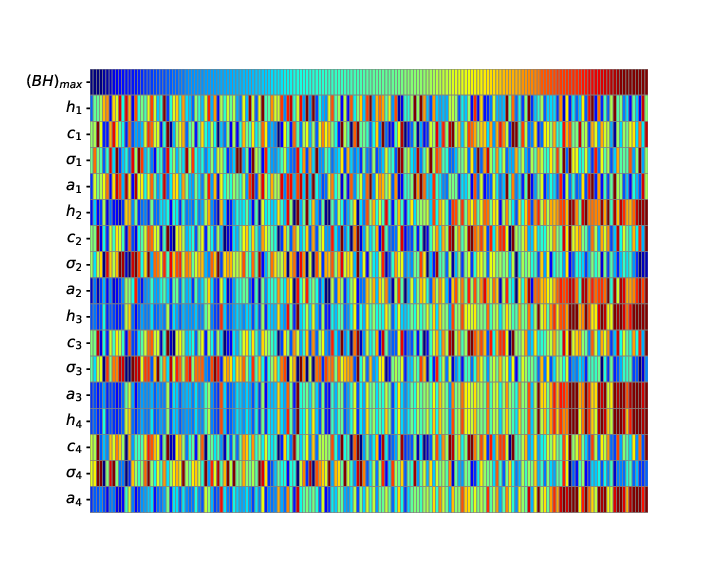}
    \caption{ML-ready XRD peak feature table built on our framework. The x-axis denotes the index of data. The suffix of the feature label denotes the index of peaks. The heatmap was sorted by $(BH)_{max}$ values.}
    \label{fig:feature_heatmap}
\end{figure}

We also considered the peak area $a_k$ as a peak feature. In this study, we denoted the feature vector for material properties as $\bm{z}=\{h_k, c_k, \sigma_k, a_k | k=1,\ldots,\mathrm{K}=4\}$. Figure \ref{fig:feature_heatmap} shows the heatmap of the feature vector standardized at each feature. The x- and y-axes show the index of XRD data and label of a feature, respectively. The suffix of the feature label denotes the peak index. The heatmap was sorted by $(BH)_{max}$ values. In this figure, warm colors exhibited a larger value and cold colors had a smaller value. This figure shows the ML-ready table data that we can input into an ML library. In the next sub-section, we present the results of Bayesian linear regression for $(BH)_{max}$ using peak features.


\subsection{Bayesian linear regression with feature selection}\label{sec:reg} 
\quad This sub-section shows the prediction of the magnetic property $(BH)_{max}$ using the peak features. In particular, our method automatically selected the peak feature needed to perform predictions using Bayesian estimation. We described the Bayesian linear regression model in Appendix \ref{sec:model:blr}. In this study, we set the variance $\lambda^{-1}$ to 0.05. Further, we used the replica exchange Monte Carlo (REMC) method in sub-section \ref{sec:algorithm} to sample from the poster distribution $p(\bm{g}, \bm{w} | \bm{t}, \bm{z})$. As a result, we estimated the optimal noise level (the optimal inverse temperature) using Bayesian free energy.

\begin{figure}
    \centering
    \includegraphics[width=\linewidth]{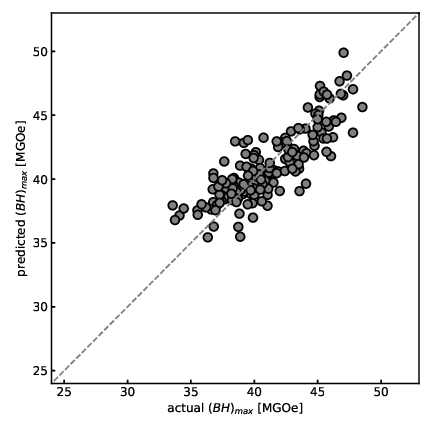}
    \caption{Results of Bayesian linear regression with feature selection. XRD peak features were used as explanatory variables.}
    \label{fig:predict}
\end{figure}

Figure \ref{fig:predict} shows the prediction results of the magnetic property $(BH)_{max}$ using the peak features. As shown in Figure \ref{fig:predict}, the regression results were reasonably good. This good regression result was achieved with only four features $\{h_1, a_1,h_3, a_3\}$ estimated by feature selection using our method. The feature subset $\{h_1, a_1,h_3, a_3\}$ indicates the peak area and height of peaks 1 and peak 2, respectively.

\begin{figure}
    \centering
    \includegraphics[width=\linewidth]{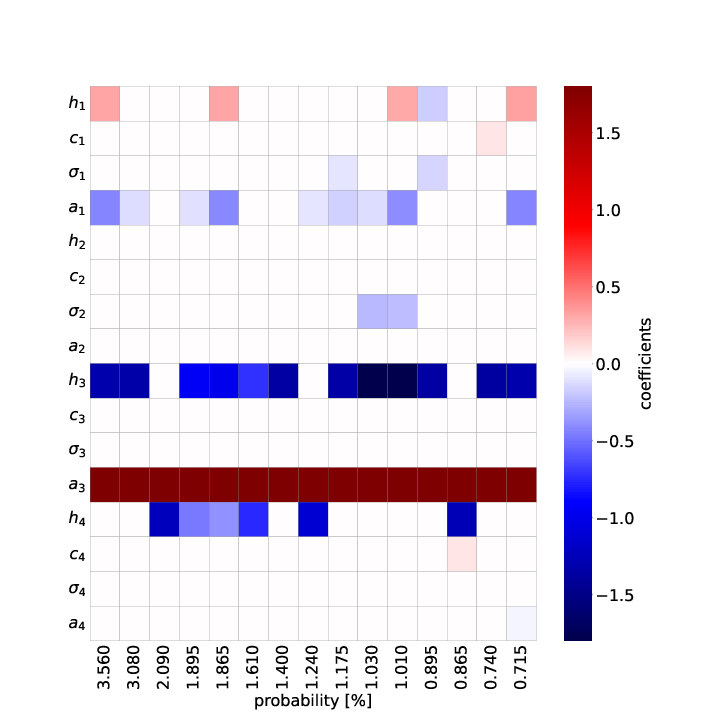}
    \caption{Results of feature selection by Bayesian linear regression. The parameter $a_3$ are suggested as the most possible from the regression.}
    \label{fig:log_c_00}
\end{figure}

Figure \ref{fig:log_c_00} shows the ranking of peak feature combinations. The x- and y-axes denote the feature labels and probability, respectively. The color indicates the estimated weight coefficients for each combination. This figure presents the rankings from 1–15. The combination on the left had a higher ranking. The regression results for the leftmost combination on are same as those shown in Figure \ref{fig:predict}. As shown in Figure \ref{fig:log_c_00}, the feature $h_3$, $a_3$ is an important feature that is essential for prediction because the features $h_3$ and $a_3$ were frequently used in higher rankings. Similarly, the features $h_1$ and $a_1$ are also considered as comparatively important features.

Comparison with the reference peak list of the $\mathrm{ Nd_{2}Fe_{14}B }$ phase, the main phase of the Neodymium magnet, shows that peaks 1 and 3 correspond to the c-planes of \textit{006} and \textit{004}, that is the axis of easy magnetization. Then we considered the weight coefficients of the \textit{004} peak area and height, which were the most important features to predict $(BH)_{max}$. The sign coefficients of the area and height were positive (red) and negative (blue), respectively, indicating that the low peak height and large peak area are expected for high $(BH)_{max}$. This suggests that decreasing crystallite size improves $(BH)_{max}$. Thus, our method provided highly interpretative results because it extracted the peak features and selected a few peak features for predicting $(BH)_{max}$.

\section{Limitation and scope}\label{sec:limit}
\quad Our framework calculated the features by focusing on ROIs with large peak intensities to accelerate the process and eliminate the intentional genre of an analyst. Therefore, our framework did not consider small peaks of intensity. Thus, the method cannot extract important features if small amounts of impurities contribute significantly to the material properties.

The proposed method required a certain number of ROIs to be the focus of attention. The larger the number of ROIs considered, the more likely it is to obtain important features, although this increases the computational complexity of feature extraction. It is recommended to set the number of ROIs at an acceptable computational cost.

\section{Conclusion}\label{sec:conclusion}
\quad This study established a methodology for fast and robust creation of ML-ready peak feature tables from XRD data. Our framework used Bayesian estimation to perform peak feature extraction from the XRD data and regression analysis with feature selection on the material properties using the peak features. Bayesian estimation facilitated discussions about the reliability (Bayesian posterior distribution) of the extracted features and regression results. In addition, our method focused only on peak tops in local ROIs for fast extraction of peak features. As a result, we could quickly obtain the main peak features and construct ML-ready feature tables from the XRD measurement data.

We applied our framework to XRD data of Neodymium magnets. We extracted the peak features with Bayesian posterior distribution from 176 XRD datapoints in about 180 min even though XRD data was analyzed in sequence. Therefore, our method enabled us to obtain peak features with confidence intervals in approximately 1 min. The magnetic properties were subjected to Bayesian linear regression using the extracted peak features and reasonably good regression results were obtained. The peak height and peak area of the \textit{004} and \textit{006} peaks were important as the necessary features, which were reasonable and interpretable.

\section*{Acknowledgment}
This work was supported by MEXT Program: Data Creation and Utilization-Type Material Research and Development Project Grant Number JPMXP1122715503, and a Grant-in-Aid for Scientific Research (KAKENHI Grants No. 19H05819).

\bibliographystyle{unsrt} 
\bibliography{myref} %

\appendix

\section{Method}\label{sec:method}

\subsection{Model | Fitting model for peak features}\label{sec:model:peak}
\quad We consider the observation process of the XRD data. The XRD data $y_n \in \mathbb{N}$ were observed from a Poisson distribution $\mathcal{P}(y_n | f_n)$ with the profile function $f_n = f(x_n;\Theta)$ as expected value:
\begin{eqnarray}
    y_n \sim \mathcal{P}( y_n | f_n ) = \frac{ f_n^{y_n} e^{-f_n} }{ y_n! }.
\end{eqnarray}

The profile function $f(x_n;\Theta)$ can be represented by linear sum of the peak signal $S(x_n;\Theta)$ and the background $B(x_n)$:
\begin{eqnarray}
        f(x_n;\Theta) = S(x_n;\Theta) + B(x_n),
\end{eqnarray}
where $(x_n, y_n)$ denotes the measured data points, the function $S(x_n;\Theta)$ denotes the peak signal, the function $B(x_n)$ denotes the background, and $\Theta$ is the profile parameter set.

The peak signal $S(x_n;\Theta)$ can be represented by a linear sum of the peak functions, which is the Voigt function $V(x_n)$ \cite{1987convVoigt}:
\begin{eqnarray}
    S(x_n;\Theta) &=& \sum_{k=1}^{\mathrm{K}}{h_k V(x_n;c_k, \sigma_k)},\\
    \text{where} \:\: V(x_n;c_k, \sigma_k) &=& \int_{-\infty}^{\infty} G(x';c_k, \sigma_k)L(x_n-x';c_k, \sigma_k)dx'.
\end{eqnarray}

The integer value $\mathrm{K}$ is the number of peaks, and the parameter set $\Theta$ is $\Theta=\{h_k, c_k, \sigma_k\}_{k=1}^{\mathrm{K}}$. The functions $G(x_n)$ and $L(x_n)$ are the Gaussian and Lorentzian functions, respectively. Although Gaussian and Lorentzian widths are often defined with different parameters, they are defined with the same parameters for simplicity in this study. The estimated parameter set $\hat{\Theta}$ corresponded to the peak features for predicting material properties.

We present a set of prior distributions as follows:
\begin{eqnarray*}
    h_k &\sim& \mathcal{G}\left( h_k | \eta=\eta(\mu_h, \sigma_h), \theta=\theta(\mu_h, \sigma_h) \right),\\
    c_k &\sim& \mathcal{N}(c_k | \mu=\mu_c, \sigma=\sigma_c),\\
    \sigma_k &\sim& \mathcal{G}\left( \sigma_k | \eta=\eta(\mu_{\sigma}, \sigma_{\sigma}), \theta=\theta(\mu_{\sigma}, \sigma_{\sigma}) \right),
\end{eqnarray*}
where the probability distributuin $\mathcal{G}(\eta, \theta)$ and $\mathcal{N}(\mu, \sigma)$ are the Gamma and the Gaussian distributions, respectively. The parameters $\eta$ and $\theta$ are shape and scale parameter of the Gamma distribution, respectively. Here, the functions $\theta(\mu, \sigma)$ and $\eta(\mu, \sigma)$ calculated the shape and scale parameter of the Gamma distribution from the mode $\mu$ and 
the variance $\sigma^2$, respectively. The functions $\theta(\mu, \sigma)$, $\eta(\mu, \sigma)$ can be expressed as follows:
\begin{eqnarray*}
    \theta(\mu, \sigma) &=& \frac{1}{2} \left( -\mu + \sqrt{ \mu^2 + (2\sigma)^2 } \right),\\
    \eta(\mu, \sigma) &=& 1 + \frac{\mu}{\theta}.
\end{eqnarray*}
Note that this variance $\sigma^2$ and peak width $\sigma_k$ have different meaning. 

As the peak height and width have positive values: $h_k \in \mathbb{R}^{+}$ and $\sigma_k \in \mathbb{R}^{+}$, we set the Gamma distribution to a prior distribution of $h_k$ and $\sigma_k$. We set the position (angle) and intensity of maximum value in the ROI to $\mu_c$ and $\mu_h$. We estimated the background $B(x_n)$ using pybaselines \cite{pybaselines} before peak extraction. 

\subsection{Model | Bayesian linear regression with feature selection}\label{sec:model:blr}
\quad The linear regression model with feature selection is represented by an output $t_m$, an input (features) vector $\bm{z}_m \in \mathbb{R}^{D}$, indicator $\bm{g} \in \{0, 1\}^{D}$ \cite{2021JampOkajima}, the weight coefficients $\bm{w} \in \mathbb{R}^{D}$, and statistical noise $\epsilon_m$ as follows:
\begin{eqnarray}
    t_m = (\bm{g} \circ \bm{w})^\top \bm{z}_m + \epsilon_m,
\end{eqnarray}
where the operation $\circ$ denotes the Hadamard product. In this study, statistical noise $\epsilon_m$ followed a Gaussian distribution with variance $\lambda^{-1}$: $\epsilon_m \sim \mathcal{N}(\epsilon_m|0, \lambda^{-1})$.

Therefore, the probability distribution of the output $t_m$ is represented as follows:
\begin{equation}
    p(t_m|\bm{g}, \bm{w}, \bm{z}_m) = \mathcal{N}( t_m| (\bm{g} \circ \bm{w})^\top \bm{z}_m, \lambda^{-1}).
\end{equation}
The prior distribution of the weight coefficients $p(\bm{w})$ was set to an uninformative broad distribution. The prior distribution of the indicator $p(\bm{g})$ is the Bernoulli distribution with the Bernoulli distribution with the expected value of 0.5.

The poster distribution of parameters $\bm{w}$, $\bm{g}$ is represented as follows:
\begin{eqnarray}
    p(\bm{w}, \bm{g} | \bm{z}, \bm{t}) &\propto& p(\bm{w})p(\bm{g}) \prod_{m=1}^{M}{p(t_m|\bm{g}, \bm{w}, \bm{z}_m)},\\
                                       &\propto& \prod_{m=1}^{M}{\mathcal{N}( t_m| (\bm{g} \circ \bm{w})^\top \bm{z}_m, \lambda^{-1})}.
\end{eqnarray}

\subsection{Algorithm | Replica Exchange Monte Carlo method}\label{sec:algorithm}
\quad We performed posterior visualization and the maximum a posteriori (MAP) estimation through sampling from the posterior distribution. A popular sampling method is the Monte Carlo (MC) method, which may be bounded by local solutions for cases when the initial value is affected or the cost function landscape is complex.

Therefore, the replica exchange Monte Carlo (REMC) method \cite{remc,2012BayesianSpectral} was used to estimate the global solution. For sampling using the REMC method, a replica was prepared with the inverse temperature $\beta$ introduced as follows:
\begin{eqnarray}
    p(\theta|\bm{y};\beta=\beta_{\tau}) &=& \exp{(-\beta_{\tau} E(\theta))}p(\theta),
\end{eqnarray}
where the function $E(\theta)$ denote a loss function, this is, a negative log likelihood with a parameter set $\theta$. The inverse temperature $\beta$ is $0 = \beta_1 < \beta_2 < \cdots < \beta_{\tau} < \beta_T = 1$. For each replica, the parameters were sampled using the Monte Carlo method.

\section{Supplement}

\subsection{Posterior distribution of peak parameters}\label{sec:sup:posterior}
\begin{figure}
    \centering
    \includegraphics[width=\linewidth]{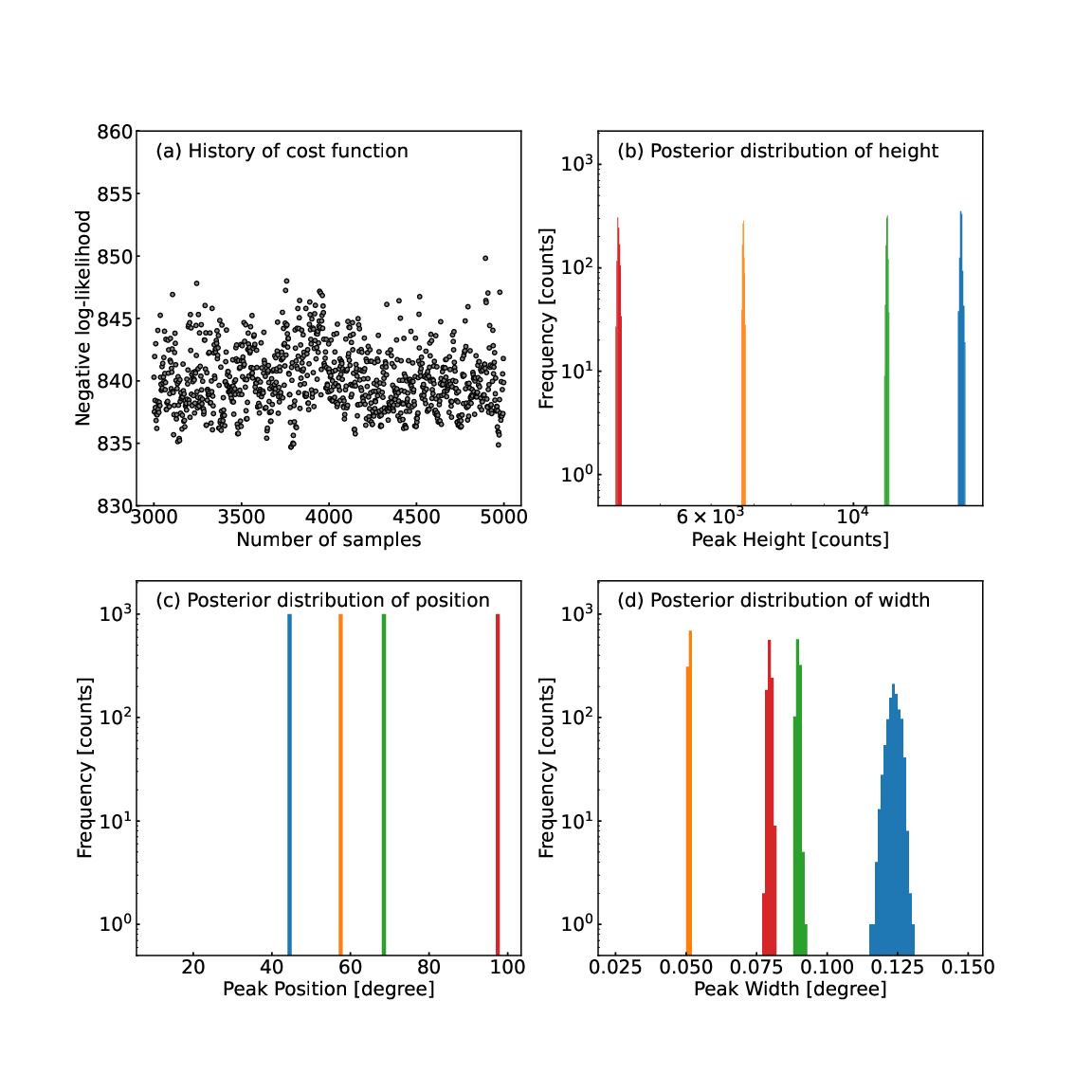}
    \caption{(a) History of cost function (the negative log-likelihood) and (b)--(d) Posterior distribution of each peak parameter: (b) peak height, (c) peak position, and (d) peak width. This figure corresponds to the Figure \ref{fig:exsample_fitting}. In this figure (b)--(d), the colors denote peak ID.}
    \label{fig:posterior}
\end{figure}

\quad Figure \ref{fig:posterior} shows the history of the cost function (the negative log-likelihood) and posterior distribution of each peak parameter: (b) peak height, (c) peak position, and (d) peak width. Figure \ref{fig:posterior} corresponds to the fitting in Figure \ref{fig:exsample_fitting}. We performed burn-in on 3000 samples, and then performed production sampling on 2000 samples. A total of 5000 MC sampling steps were performed for fitting of one spectral datapoint. To reduce the correlation between samples, we visualised the posterior distribution using the history of 1000 MC samples at one sample interval. As shown in Figure \ref{fig:exsample_fitting}(b)--(d), we believed that the posterior distributions had sufficient sharpness to use material description feature. In particular, peak positions were estimated with high precision. If the posterior distribution is very broad, we can eliminate this feature from the explanatory variables. Our method can provide a posterior distribution for discussing the reliability of the features.

\end{document}